\documentclass[superscriptaddress,citeautoscript,aip,jcp,preprint,footinbib]{revtex4-1}
\usepackage{graphicx}
\usepackage{enumitem}
\usepackage{gensymb}
\usepackage{amsmath}
\usepackage{amssymb}
\usepackage{dcolumn}
\usepackage{color}
\begin{document}

\title{The color center singlet state of oxygen vacancies in $\text{TiO}_\text{2}$}

\author{Ji Chen}
\email{ji.chen@pku.edu.cn}
\affiliation{School of Physics, Peking University, Beijing 100871, P. R. China}
\affiliation{Max Planck Institute for Solid State Research, Heisenbergstrasse 1, 70569 Stuttgart, Germany}

\author{Nikolay A. Bogdanov}
\affiliation{Max Planck Institute for Solid State Research, Heisenbergstrasse 1, 70569 Stuttgart, Germany}

\author{Denis Usvyat}
\affiliation{Institut f\"{u}r Chemie, Humboldt-Universit\"{a}t zu Berlin, Brook-Taylor-St. 2, D-12489 Berlin, Germany}

\author{Wei Fang}
\affiliation{Laboratory of Physical Chemistry, ETH Zurich, CH-8093 Zurich, Switzerland}

\author{Angelos Michaelides}
\affiliation{Department of Physics and Astronomy, London Centre for Nanotechnology, Thomas Young Centre, University College London, Gower Street, London WC1E 6BT, U.K.}
\affiliation{Max Planck Institute for Solid State Research, Heisenbergstrasse 1, 70569 Stuttgart, Germany}
\affiliation{Department of Chemistry, University of Cambridge, Lensfield Road, Cambridge CB2 1EW, United Kingdom}

\author{Ali Alavi}
\email{a.alavi@fkf.mpg.de}
\affiliation{Max Planck Institute for Solid State Research, Heisenbergstrasse 1, 70569 Stuttgart, Germany}
\affiliation{Department of Chemistry, University of Cambridge, Lensfield Road, Cambridge CB2 1EW, United Kingdom}

\begin{abstract}
Oxygen vacancies are ubiquitous in 
$\text{TiO}_\text{2}$ and play key roles in catalysis and magnetism applications.
Despite being extensively investigated, the electronic structure of oxygen vacancies in $\text{TiO}_\text{2}$ remains 
controversial both experimentally and theoretically.
Here we report a study of 
a neutral oxygen vacancy in $\text{TiO}_\text{2}$ using state-of-the-art quantum chemical electronic structure methods.
We find that the ground state is a color center singlet state in both the rutile and the anatase phase of $\text{TiO}_\text{2}$. 
Specifically, embedded CCSD(T) calculations find, for an oxygen vacancy in rutile, that the lowest triplet state energy is 0.6 eV above the singlet state, and in anatase the triplet state energy is higher by 1.4 eV. 
Our study provides fresh insights on the electronic structure of the oxygen vacancy in $\text{TiO}_\text{2}$, clarifying 
earlier controversies and 
potentially
inspiring future studies of defects with correlated wave function theories.
\end{abstract}

\maketitle

\section{Introduction}

Titanium oxides are widely known in clean energy technologies such as solar water splitting
and applications 
due to
defect magnetism \cite{diebold_surface_2003,thompson_surface_2006,fujishima_tio2_2008, venkatesan_thin_2004,yoon_oxygen-defect-induced_2006,hong_room_2006}.  
Studies from different perspectives have all pointed out that oxygen vacancies are particularly important in $\text{TiO}_\text{2}$, along with other defects such as interstitials and cation impurities.  \cite{henderson_surface_2011,pang_structure_2013}.
%
%
Pristine $\text{TiO}_\text{2}$ has empty $3d$ orbitals in the conduction band, and a neutral oxygen vacancy effectively dopes two
excess electrons.
Since, in $\text{TiO}_\text{2}$, the oxygen site is three-fold coordinated with Ti, the two excess electrons are
likely to occupy the three nearest Ti sites and the vacancy, forming a localized defect.
%
However, with such a defect embedded in the crystalline lattice of a solid, the electron correlation problem is
far from trivial, and the resulting electronic structure of the neutral oxygen vacancy may require a high-level quantum chemical treatment for a reliable description.

%
Experimental studies have advanced from exploring the macroscopic behaviors of defects towards
precise measurements on isolated defects focused on elucidating their
fundamental electronic structure.
Deep band-gap states at about 1 eV below the conduction band were identified in electron energy loss spectroscopy, photoelectron spectroscopy, and scanning tunneling microscopy studies \cite{yim_oxygen_2010, henderson_insights_2003, setvin_direct_2014}.
Shallow electron donor defect states were observed in infrared spectroscopy and 
electron paramagnetic resonance (EPR) studies \cite{panayotov_infrared_2012, yang_photoinduced_2009,brandao_2009,yang_intrinsic_2013, brant_triplet_2014}.
EPR experiments have detected both the triplet (S=1) and the doublet (S=1/2) states
in $\text{TiO}_\text{2}$ with oxygen vacancies, but
the relative stability of different spin states has not been determined 
\cite{yang_photoinduced_2009,brandao_2009,yang_intrinsic_2013, brant_triplet_2014}. 
%

Complementary to experimental measurements, electronic structure calculations have 
been established as useful tools to 
investigate the roles of oxygen 
vacancies \cite{hussain_structure_2017,selcuk_facet-dependent_2016, schaub_oxygen_2001, yoon_oxygen-defect-induced_2006,hong_room_2006}.
In particular, density functional theory (DFT) studies have reproduced reasonably well the band gap and the defect levels of $\text{TiO}_\text{2}$, and 
revealed the activity of surface vacancies for water splitting \cite{mattioli_ab_2008,valentin_reduced_2009,janotti_hybrid_2010,morgan_dft_2007,stausholm-moller_DFT_2010,morgan_intrinsic_2010,deak_quantitative_2012,janotti_dual_2013,bjorheim_defect_2013}.
However, DFT with commonly used exchange correlation functionals is not free of errors and the electronic structure of the oxygen vacancy in $\text{TiO}_\text{2}$
varies from one study to another.
Regarding the spin state of the neutral oxygen vacancy in rutile, Mattioli et al. found that the triplet is more stable than the singlet \cite{mattioli_ab_2008,lee_calculation_2012, stausholm-moller_DFT_2010}; Janotti et al. identified the singlet as the ground state \cite{janotti_hybrid_2010}; De\'ak et al. found a near degeneracy of the singlet and the triplet \cite{deak_quantitative_2012}; and many studies have pointed to uncorrelated polaronic defects of $\text{Ti}^{3+}$ (S=1/2) \cite{deak_quantitative_2012,janotti_dual_2013, divalentin_reduced_2009, deskins_electron_2007}. 
Moreover, in these studies the spin charge density distributions around the vacancy were quite different, with some cases showing occupation at the nearest Ti sites and others showing charge density at the vacancy \cite{mattioli_ab_2008, janotti_hybrid_2010, deak_quantitative_2012, malashevich_first_2014}.
The discrepancies may stem from the different choice of the exchange correlation functional in DFT, but
it is also possible that none of the available functionals is reliable for such a problem.
Many body perturbation theory (specifically the GW approximation) has also been used 
to correct the predictions of the stability of defect states
in $\text{TiO}_\text{2}$ \cite{malashevich_first_2014, Chen_origin_2018},
but the spin states of oxygen vacancies were not discussed.

Overall, a benchmark of the electronic spin states of oxygen vacancies at a higher level of theory
is desired. 
Quantum Monte Carlo (QMC) and wave function theory (WFT) methods have been successfully applied to molecules and model systems to tackle electron correlation problems.
More recently, these methods have also been used to study solids
 \cite{booth_towards_2013,libisch_embedded_2014,michealides_preface_2015,denis-to-submit,kubas_surface_2016,Bogdanov_2016,tsatsoulis_comparison_2017}.
Although there are challenges to be overcome with the application of such methods to solids, the outcome of such highly accurate, parameter-free, calculations provides insightful information of the true electronic structure of the systems studied.

%
In this study we investigate the electronic structure of a
neutral oxygen vacancy in $\text{TiO}_\text{2}$ using correlated wave function theories.
With the help of preliminary full configuration interaction quantum Monte Carlo (FCIQMC) calculations \cite{booth_fermion_2009, cleland_survival_2010}, we find that the nature of both the singlet and the triplet states are predominantly single-reference. 
Coupled cluster theory calculations are then carried out to obtain the energy of the states, from which we find that
the singlet state is more stable than the triplet state both in the rutile and the anatase phases.
With accurate benchmarks, we discuss the performance of DFT and find that it underestimates the stability of the singlet state, which helps to resolve existing controversies.

\section{Methods}

In this study we applied an embedded cluster scheme using the 
ChemShell package \cite{metz_chemshell_2014}.
Figure \ref{figure_emodel}a shows the embedding model for rutile, which is composed of three shells.
The first shell contains 3 Ti atoms and their nearest oxygen neighbors (14 O atoms for rutile and 13 O atoms for anatase), and was treated at the all-electron quantum level.
We use the cc-pVDZ basis set \cite{MOLPRO-BASIS-CC,CCbasis_O} for all atoms in the all-electron shell. Oxygen basis functions were 
also placed at the site of the vacancy. 
Farther Ti atoms (19 for rutile and 16 for anatase) surrounding the all-electron oxygen were treated with the effective core 
potential (ECP) of Wedig et al. \cite{wedig_basis} that incorporates 18 electrons in the core augmented with one $s$ function. 
A larger cluster of rutile (22 all-electron Ti atoms, 76 all-electron O atoms, and 51 ECP Ti atoms) was used to test the convergence of our calculations with respect to the size of the cluster.
Outside of the quantum-mechanical cluster, a spherical shell with radius 20 \AA~ of formal point charges at crystallographic positions was used in conjunction
with additional charges fitted to reproduce the correct Madelung electrostatic potential in the quantum region.
Within the embedded cluster we performed 
FCIQMC \cite{booth_fermion_2009, cleland_survival_2010} calculations with the NECI code \cite{booth_linear_2014, neci_paper} and WFT-based calculations using MOLPRO \cite{molpro-wires}.
Restricted Hartree Fock (HF) and restricted open shell HF self consistent field calculations were used for the singlet and the triplet, respectively.
The correlated WFT methods used include second order M{\o}ller-Plesset perturbation theory (MP2), coupled cluster with singles and doubles (CCSD), and
CCSD with perturbative triples (CCSD(T))\cite{watts_coupled_1993, knowles_coupled_1993}.
%
Geometry optimizations were performed using a Python interface to MOLPRO.
DFT calculations were performed with the 
PBE0 \cite{perdew_rationale_1996, Adamo_toward_1999} and
B3LYP \cite{devlin_ab_1995} exchange correlation functionals.
%
%
%
More details and additional computational tests are given in the supporting information (SI).

\section{results and discussions}

\begin{figure}[htb]
\begin{center}
\includegraphics[width=8cm]{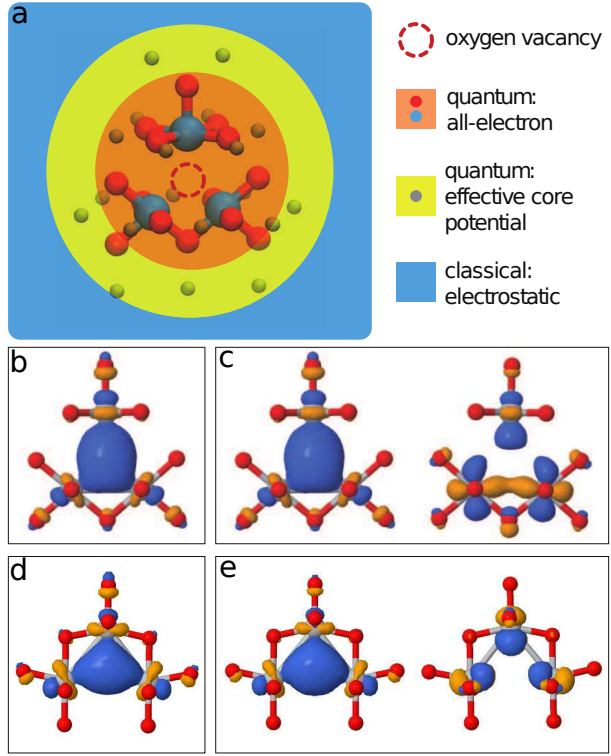}
\end{center}
\caption{(a) Embedding model of the oxygen vacancy in rutile, showing three shells from inside out, namely 
the all-electron quantum cluster, the effective core potential buffer layer, and the classical electrostatic regime.
(b) The doubly occupied frontier orbital of the singlet state in rutile from DFT-PBE0.
(c) The frontier orbitals of the triplet state in rutile.
(d) The cluster structure and the doubly occupied frontier orbital for the anatase phase.
(e) The frontier orbitals of the triplet state in anatase.
}
\label{figure_emodel}
\end{figure}

%
Two excess electrons induced by a neutral oxygen vacancy may form a spin singlet state or a triplet state.
%
Fig. \ref{figure_emodel}b-c and d-e shows the DFT-calculated frontier orbitals of the lowest energy singlet and triplet states in the rutile and the anatase clusters, respectively.
Although similar orbitals are found in the HF calculations,
there is a major discrepancy between DFT and HF state energies.
DFT predicts that the ground state is the singlet state, whereas HF strongly favors the triplet.
Before getting into a quantitative assessment of the states, we 
note that both the singlet and the triplet involve a so-called color center orbital, featuring electron localization in the center of the vacancy.
Specifically, in the singlet state the color center orbital is doubly occupied, while in the triplet state it is singly occupied together with a delocalized molecular orbital formed by $\text{Ti}_{3d}$ orbitals around the vacancy.
%
The frontier orbitals of the singlet and triplet states suggest that the electronic structure of oxygen vacancy does not feature a covalent picture between Ti and coordinating O atoms, but $\text{Ti}^{3+}$ around the vacancy are connected by the color center state (Fig. S4).

\begin{figure}[htb]
\begin{center}
\includegraphics[width=8cm]{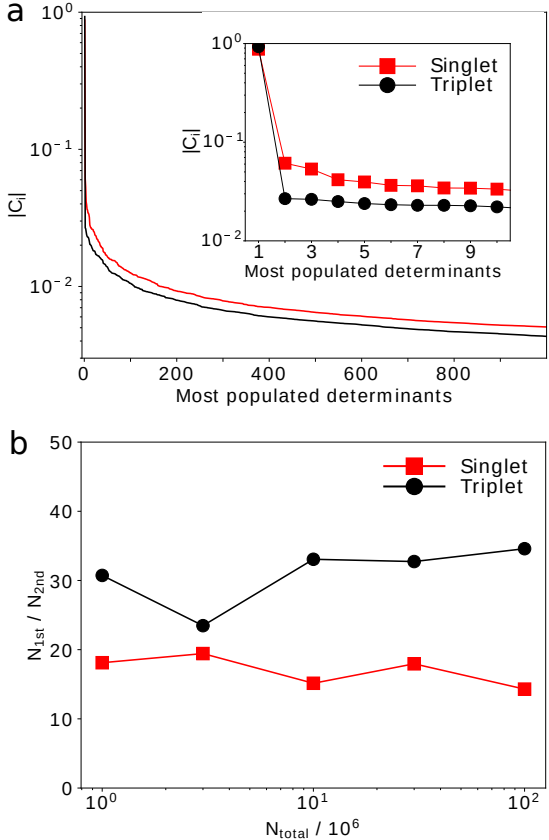}
\end{center}
\caption{
FCIQMC results of the singlet and the triplet states.
(a) The weight of the most populated determinants as excitations from the reference with 100 million walkers in total.
$|C_i|$ 
is defined as the $|N_i|/ \sqrt{\sum_i {N_i^2}}$, where $N_i$ is the population number at $i$th determinant and the sum over $i$ runs over all initiators.
The most populated determinant is the reference determinant, which has a weight of 0.88 and 0.93 for the singlet and the triplet, respectively.
(b) The relative ratio of the first and the second most populated determinant as a function of the total number of walkers.
}
\label{figure-fciqmc}
\end{figure}

To reveal the electronic structure of the singlet and the triplet states,
we carried out FCIQMC simulations \cite{booth_fermion_2009}.
A full configuration interaction (FCI) wave function is expressed as 
$\Phi_0^{FCI}=\sum_i C_i |D_i \rangle$, namely a linear expansion over all possible Slater determinants $|D_i\rangle$.
With FCIQMC the $C_i$ coefficients are optimized stochastically, gradually approaching the exact wave function.
In addition to determination of the state energy, FCIQMC makes it possible to reveal weights of individual determinants in the wavefunction.
If a clear pattern can be seen in the expansion, even before full convergence is achieved in the energetic sense, 
it may be possible to suggest a deterministic WFT method suitable for the problem.
We performed FCIQMC simulations separately for the singlet and the triplet states.
The corresponding weights of the 1000 leading (i.e. most populated) determinants are plotted in Fig. \ref{figure-fciqmc}.
The inset shows
the ratio of the second most populated determinant to the reference determinant as a function of the number of FCIQMC walkers up to
100 million.
It is clear that the $C_i$ coefficient of the reference determinant is at least approximately 0.9, for both the singlet and the triplet states.
The second most populated determinant has a population that is 20-30 times smaller than the reference.
These calculations unambiguously reveal that both the singlet and the triplet states have single-reference character.

\begin{figure}[htb]
\begin{center}
\includegraphics[width=8cm]{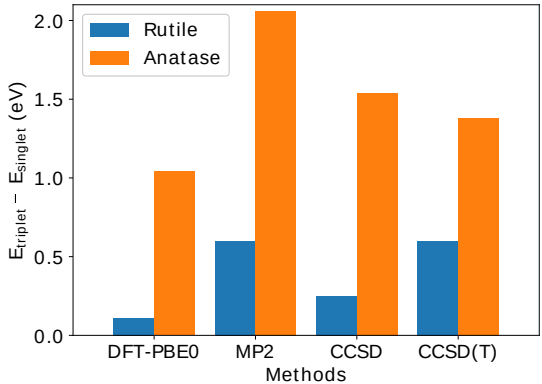}
\end{center}
\caption{
The triplet-singlet energy difference calculated using different methods. Positive values indicate that the singlet is more stable.
}
\label{figure-cluster}
\end{figure}

We now turn to quantitative computations to determine which state is the ground state and 
how stable it is.
Although the size of our system (114 electrons in 221 orbitals) makes the full convergence of the FCIQMC energies prohibitively expensive at present, 
the demonstration that the two states are both single-reference allows us to seek other appropriate WFTs to compute the energies.
Among WFTs, the coupled cluster theory is particularly powerful for single reference systems, especially at the CCSD(T) level, where
singles and double excitation amplitudes are fully solved for, and the effects of triple excitations are perturbatively treated.
In Fig. \ref{figure-cluster} we plot the energy difference between the triplet state and the singlet state.
CCSD(T) gives an energy difference of approximately 0.6 eV for rutile and 1.4 eV for anatase, both in favor of the singlet being the ground state.

%
Upon comparing CCSD(T) and CCSD we find that the
perturbative triples correct the CCSD results in a different manner for the two phases.
For rutile, CCSD underestimates the triplet-singlet energy difference,
whereas for anatase it slightly overestimates the result.
A similar discrepancy is found for MP2, which reproduces the CCSD(T) result for rutile quite well, but severely overestimates the energy difference for anatase.
We also find that DFT severely underestimates the stability of the singlet state, and in particular for the rutile phase the singlet is only marginally more stable than the triplet by approximately 0.1 eV.
In the SI, we show calculations with several other methods.
Apart from the HF calculations in which electron correlations are completely neglected, all other methods agree qualitatively that the singlet state is the ground state. 

\begin{figure}[htb]
\begin{center}
\includegraphics[width=8cm]{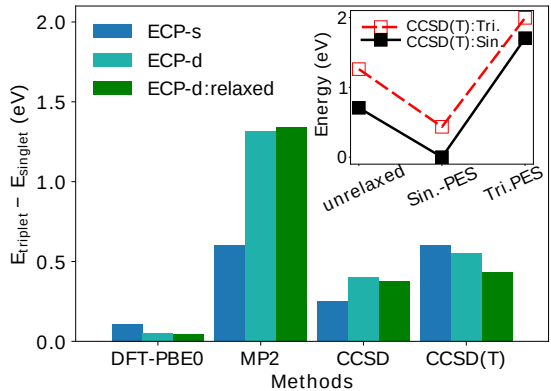}
\end{center}
\caption{
The triplet-singlet energy difference with different basis on ECP sites and different geometries for rutile.
Inset is the relative energy of the triplet and the singlet states calculated using CCSD(T) at different structures, including the unrelaxed structure and the structures relaxed along the singlet and the triplet potential energy surfaces using DFT-PBE0.
}
\label{figure-geo}
\end{figure}

Because of the relatively small triplet-singlet gap calculated in rutile, we now discuss the sensitivity of the calculations to various settings with rutile.
Overall, we find that the CCSD(T) calculations do not depend sensitively on the computational settings.
Nevertheless, we focus on two particularly important aspects summarized in Fig. \ref{figure-geo}.
Previous studies have shown that the delocalized Ti 3$d$ orbitals are particular important for such defect states. \cite{janotti_hybrid_2010,spreafico_nature_2014,chen_small_2020}
But the second shell of our system is described with large-core Ti ECPs with only one $s$ function. 
Thus, we analyze the effect of having 3$d$ basis functions at these surrounding Ti sites.
Two sets of calculations using exclusively one $s$ (referred to as ECP-s) and two $d$ functions (referred to as ECP-d) were performed (SI).
%
It turns out that the basis functions on top of the ECPs have only a negligible impact on our best CCSD(T) results, although it significantly changes the MP2 numbers.
We have also examined the impact of structure relaxation on our results.
To this end, we relax the two inner shells of the cluster separately along the singlet and the triplet potential energy surface (PES) using DFT (Fig. \ref{figure-geo} inset).
%
%
We find that after such structure relaxations, the energies of the lowest singlet and the lowest triplet maintain a gap that is only slightly smaller than the unrelaxed structure.
The singlet state remains as the ground state.

We checked that our conclusions are converged with respect to the size of the cluster by carrying out DFT calculations using a larger $\text{Ti}_{73}\text{O}_{76}$ embedded cluster (SI).
These calculations are found to be consistent with the $\text{Ti}_{22}\text{O}_{14}$ (including 19 Ti atoms treated with ECPs) results presented above.
The rapid convergence with respect to the cluster size is likely a consequence of the local nature of both the singlet and triplet defect states. A similar observation was reported in a study of semiconductor band gap using the similarity transformed equation of motion coupled cluster. \cite{dittmer_accurate_2019}
Another study showed that DFT and CCSD(T) are consistent in converging the adsorption energies on a $\text{TiO}_\text{2}$ surface with respect to the cluster size, which lends support to the conclusion that our CCSD(T) results are insensitive to the choice of cluster. \cite{kubas_surface_2016}.
To further test the validity of our findings, the classical point-charge embedding used in this
study is compared with an advanced HF embedding scheme (Fig. S6).
To obtain the HF embedding, we first performed periodic restricted closed shell HF calculations on the supercell. 
The HF Bloch wave functions were localized to Wannier functions \cite{zicovich_general_2001}, 
which were further processed to extract the orbitals of the fragment near the vacancy in real space \cite{denis-to-submit}.
To compare directly with the classical embedding described above we have selected the  same cluster size.  
The HF embedding results clearly show that the two embedding schemes are in line with each other.
Additional tests on e.g. the number of correlated electrons and the basis set are included in the SI.
Interestingly, we observe higher sensitivity of the results on computational settings for the lower level of WFT theories such as MP2 and CCSD, thus much care needs to be taken when calculations are performed with these methods.
For DFT, it appears that calculations with PBE0 and B3LYP are not very sensitive to the computational settings, including variation of the fraction of the exact exchange (SI).
However, even though PBE0 and B3LYP give consistent predictions, both severely overestimate the relative stability of the triplet state.
Moreover, when using the local density approximation (LDA) or a generalized gradient approximation (GGA) functional, severe convergence difficulties are found for the triplet state.
Therefore, it is clear that the standard approximations of DFT have issues in treating such problems.
Considering that the energy difference between the singlet and the triplet becomes almost negligible when the structure is locally optimized, this may explain the discrepancies in the literature \cite{mattioli_ab_2008,lee_calculation_2012,stausholm-moller_DFT_2010,janotti_hybrid_2010,deak_quantitative_2012,janotti_dual_2013, divalentin_reduced_2009, deskins_electron_2007}. 

Before we conclude, it is worth noting that in $\text{TiO}_\text{2}$ the defect states associated with oxygen vacancies are often discussed as polarons in experiments and corresponding DFT studies \cite{setvin_direct_2014,reticcioli_polaron_2017,reticcioli_interplay_2019,yim_engineering_2016,yim_visualization_2018,kick_towards_2019,guo_probing_2020,chen_small_2020}.
Although we have not directly modelled the polarons, which are sensitive for example to surfaces, and additionally would require long range electron-phonon couplings to be considered, 
it is likely that the polaronic states are also incorrectly described in DFT calculations
using standard exchange correlation functionals.
An important feature of polarons is the localization of electrons at Ti sites, and in the case of small polarons, local lattice distortions are induced by the change of electrostatics.
The localization of electrons is similar to what happens in the triplet state of our simulations.
Previous DFT studies have shown that LDA and GGA functionals are not sufficient to model polarons,
which is consistent with our finding that the triplet state does not converge.
As a result, most DFT studies have employed the so-called LDA+U method or hybrid functionals to treat polarons.
Our results show that although hybrid functionals can converge a localized electronic state, they tend to over-stabilize them.
Therefore, reliable modeling of the true polaronic states has not been achieved, and require further developments in the future to 
apply more accurate electronic structure theories with periodic methods and advanced embedding schemes \cite{booth_towards_2013,denis-to-submit,libisch_embedded_2014,kubas_surface_2016,kick_towards_2019}.
In addition, quantitative calculations of the electronic states are desired on geometries relaxed with accurate electronic structure methods.
Moreover, previous DFT calculations of the charge-state transition levels have suggested that the excited states of neutral oxygen vacancy are likely to be very close to or above the conduction band minimum \cite{janotti_hybrid_2010}. On the one hand, this raises a question about whether the excited triplet state is experimentally detectable. On the other hand, since the conduction band of $\text{TiO}_\text{2}$ and the triplet defect state resemble each other in occupying the empty 3d orbitals of Ti, our study indicates that previous predictions of the charge-state transition level may also be modified when accurate electronic structure calculations are applicable.

\section{Conclusions}
In summary, we have reported an \textit{ab initio} study of the neutral oxygen vacancy in $\text{TiO}_\text{2}$.
FCIQMC calculations show that the ground state singlet is dominated by a closed shell 
color center.
Further WFT and DFT calculations find consistently the singlet to be the ground state.
In particular,
CCSD(T) calculations predict that the singlet is 
more stable than the triplet state by 0.6 eV in rutile and by 1.4 eV in anatase.
%
%
%
%
So far the color center singlet spin state has not been confirmed experimentally, and our prediction that it is
a rather stable ground state suggests that its properties should be observable in future experiments.
%
%
In addition, the fact that quantum chemical methodologies including CCSD(T) predict the ground state to be a singlet, with a singlet-triplet gap which is rather sensitive to the crystal structure, may lead to new theoretical models for understanding the defect magnetism and spin-sensitive catalysis in different $\text{TiO}_\text{2}$ phases. 
%
%
Our results highlight the importance of color center orbitals and singlet states, and the question of whether their importance has been underestimated in other oxides remains for future studies.

\section*{Supplementary Material}
See supplementary material for additional results using different methods and settings, convergence tests and analyses.

\section*{Data availability}
The data that support the findings of this study are available from the corresponding author upon reasonable request.

\section*{Acknowledgements}
The authors thank Daniel Kats and Jeremy O. Richardson for sharing their codes and helpful discussions.
This work was supported by the National Key R\&D Program of China under Grant No.2016YFA030091, the National Natural Science Foundation of China under Grant No. 11974024. 
J.C. and A.M. are grateful to the Alexander von Humboldt Foundation for a post doctoral
research fellowship and a Bessel Research Award, respectively.
D.U. thanks the Deutsche Forschungsgemeinschaft for financial support (Grant US103/1-2).
We thank the TianHe-1A supercomputer, the High Performance Computing Platform of Peking University, the Platform for Data Driven Computational Materials Discovery of the Songshan Lake Materials Lab, and the Max-Planck-Gesellschaft for computational resources.

%

\end{document}